\documentclass{aa}  
\usepackage{graphicx}
\usepackage{txfonts}
\usepackage{natbib}

\def\target{V395\,Car\,\,}

\def\m2s2{m$^{2}$~s$^{-2}$} 	
\def\kms{\,km~s$^{-1}$}       	
\def\vsini{$v$\,sin\,$i$}     
\def\msol{\,M${_\odot}$}        
\def\rsol{\,R${_\odot}$}      	
\def\lsol{\,L${_\odot}$}      	
\def\angs{\,\AA}         

\begin{document}
   \title{The rotational broadening of V395\,Car -- implications on the 
   compact object's mass
   \fnmsep\thanks{Based on observations collected at the European Southern 
   Observatory, Chile, under the programme 077.D-0579A}}


   \author{T.\,Shahbaz\inst{1} \and C.A.\,Watson\inst{2}}

   \offprints{T. Shahbaz}

   \institute{Instituto de Astrofisica de Canarias, 
              c/ via Lactea s/n, 
              La Laguna, 
	      E38200,
	      Tenerife, 
	      Spain\\
              \email{tsh@iac.es}
         \and
             Department of Physics and Astronomy, 
	     University of Sheffield,
	     Hicks Building,
	     Sheffield,
	     S3 7RH,
	     England\\
             \email{c.watson@sheffield.ac.uk}
             }


 
  \abstract 
{The masses previously obtained for the X-ray binary
2S\,0921--630 inferred a compact object that was either a high-mass neutron
star or low-mass black-hole, but used a previously published value 
for the  rotational broadening (\vsini\ ) with large uncertainties .}
{We aim to determine an accurate mass for the compact object through an
improved measurement of the secondary star's projected equatorial rotational
velocity.}
{We have used UVES echelle spectroscopy to determine the \vsini\ of the
secondary star (V395\,Car) in the  low-mass X-ray binary 2S\,0921--630 by
comparison to an artificially  broadened spectral-type template star. In
addition, we have also measured \vsini\ from a single high signal-to-noise
ratio absorption line profile calculated using the method of  Least-Squares
Deconvolution (LSD). }
{We determine \vsini\ to lie between 31.3$\pm$0.5\kms to 34.7$\pm$0.5\kms
(assuming zero and continuum limb darkening, respectively) in disagreement with
previous results based on  intermediate resolution spectroscopy obtained with
the 3.6m NTT. Using our revised \vsini\ value in combination with the secondary
star's radial velocity gives a binary mass ratio of 0.281$\pm$0.034.
Furthermore, assuming a binary inclination angle of 75$^\circ$ gives a compact
object mass of 1.37$\pm$0.13\msol.}
{We find that using relatively low-resolution spectroscopy can result in
systemic uncertainties in the  measured \vsini\ values obtained using standard
methods. We suggest the use of LSD as a secondary,  reliable check of the
results as LSD allows one to directly discern the shape of the absorption line
profile. In the light of the new \vsini\ measurement, we have revised down the
compact object's mass,  such that it is now compatible with a canonical neutron
star mass.}  

   \authorrunning{Shahbaz et al.}
   \titlerunning{The rotational broadening of V395\,Car -- the 
   compact object's mass}

   \maketitle

\section{Introduction}

2S0921--630 was identified with a $\sim$16 mag blue star \citep{Li78},
V395\,Car, whose optical spectrum was characteristic of a low-mass
X-ray binary \citep{Branduardi83}.   The secondary star is thought to
be a halo object  in orbit with a K0\,III companion star; absorption
lines of the companion have been detected (\citealt{Branduardi83};
\citealt{Shahbaz99}; \citealt{Shahbaz04}; \citealt{Jonker05}).  Indeed
\target is one of those rare low-mass X-ray binaries in which the secondary is visible
despite the presence of a luminous disk. The inclination angle of
\target is relatively well constrained and must be high since partial
eclipses of the compact object and accretion disc (typical for 
accretion-disc corona sources) have been observed
at both X-ray and optical wavelengths (e.g.  \citealt{Branduardi83};
\citealt{Chevalier81}; \citealt{Mason87}).  There has been no
detection of type I X-ray bursts of pulsations, so the nature of the
compact object is unclear.

Intermediate-resolution optical spectroscopy revealed the donor to be
a K0\,III star with a rotational velocity of \vsini=64$\pm$9\kms,
contributing 25\% to the observed flux near H$\alpha$
\citep{Shahbaz99}.  Further optical spectroscopy yielded the 
the orbital period (9.0035 d) and the secondary
star's radial velocity 
which, when combined with the \vsini\
measurements (with rather large uncertainties), led to the conclusion
that  the system contains either a massive neutron star or a low-mass
black hole (\citealt{Shahbaz04}; \citealt{Jonker05}). In this paper we
present the results of UVES echelle spectroscopy  aimed at refining
the \vsini\ measurement, which is essential for an accurate determine
of the binary masses.

\begin{table}
\caption{Log of observations. The orbital ephemeris was taken from \citet{Shahbaz04}.}
\label{log}
\begin{center}
\begin{tabular}{lccc}\hline \hline
Object       & UT Date   & UTC   & Orbital phase \\
\hline
 V395\,Car   &  13/04/06 & 23:47:06 & 0.19 \\
 V395\,Car   &  22/04/06 & 01:23:09 & 0.08 \\
 V395\,Car   &  10/05/06 & 00:20:52 & 0.08 \\
 HD83155     &  10/05/06 & 01:05:56 & -    \\ \hline	 
\end{tabular}
\end{center}
\end{table}

\section{Observations and Data reduction}
\label{obs}

We obtained spectra of \target in service mode with the UV-Visual Echelle
Spectrograph (UVES) at the European Southern Observatory (ESO) Observatorio
Cerro Paranal, using the 8.2 m Very Large Telescope (VLT). A log of
observations is given in  Table\,\ref{log}. The UVES standard dichroic DIC1 was
used yielding spectra covering the spectral ranges 3800--4500\angs\ (blue),
4800--5800\angs\ (green) and 5800--6800\angs~(red). The blue spectrum is
recorded with a single CCD detector, while the red arm is covered by a mosaic
of two CCD chips, leading to a small gap in the red spectrum. Three spectra of
\target were taken using an exposure time of 2400\,s and a K0\,III spectral type 
template star
(HD83155) was also observed.  An 0.8 arcsec slit was used resulting in a
resolving power of 43,000 and an instrumental velocity resolution of 6.4\kms\
(FWHM). We used the UVES pipeline software which provides an absolute flux
calibrated  spectrum. The procedure consisted of bias subtraction,
flat-fielding, wavelength calibration using ThAr lamps and absolute flux
calibration.

\begin{figure}
\begin{center}
\includegraphics[width=5.5cm,height=8.5cm,angle=-90]{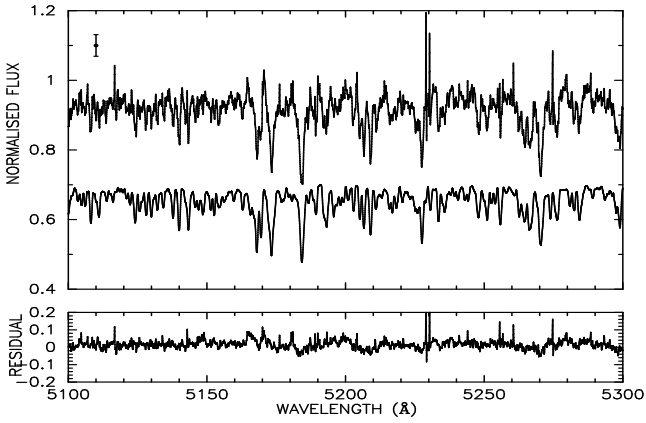}

\caption{ The small section of the green spectra of \target. In the upper
panel the top spectrum is the \target and the bottom spectrum is the
K0\,III template star rotationally broadened by 34.7\kms\ using a
continuum limb-darkened  spherical rotation profile and scaled to
match the  top spectrum. The error bars shows the typical uncertainties. There 
are a number of resolved lines in the \target spectrum that are well matched 
by the K-star.
The bottom panel shows the residual spectrum
after the optimal subtraction. }
\end{center}
\label{spectrum}
\end{figure}

\begin{figure}
\begin{center}
\includegraphics[width=5.5cm,height=8.5cm,angle=-90]{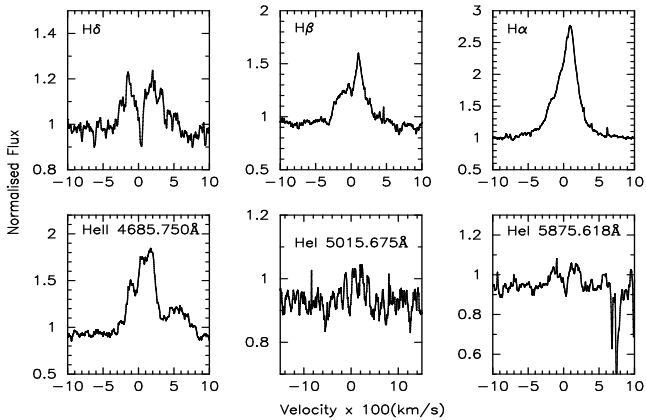}
\caption{Zooms of the most prominent emission line features.}
\end{center}
\label{features}
\end{figure}

\section{The rotational broadening}
\label{vsini}

We normalized the individual \target and template star flux calibrated  spectra
by dividing by a first-order polynomial fit, and then subtracting a high-order
spline fit to carefully selected continuum regions. This ensures that line
strength is preserved along the spectrum and is particularly important when
the absorption lines are veiled by differing amounts over a wide wavelength
range. The individual spectra were then corrected for radial velocity shifts,
determined by cross correlating the individual \target spectra with the 
template K0\,III star (using regions devoid of emission and interstellar lines), 
and then combined in order to improve the signal-to-noise ratio.  The \target
and template star spectra were then binned on the same  uniform velocity
scale. The final blue spectrum of  \target had a signal-to-noise ratio of
15 in the continuum, whereas the green and red 
spectra  of \target had a signal-to-noise ratio of 35 in the continuum (see
Figure\,\ref{spectrum}). A zoom of the significant emission line features is
shown in Figure\,\ref{features}.

The spectra of \target were taken near orbital phase 0.0 (inferior conjunction
of the secondary star) when the secondary star's contribution to the observed
flux is at its most. Although, the smearing of the spectral lines associated
with the radial velocity of the secondary star during its orbital motion is
also at its maximum, the relatively short exposures (2400 s) compared to the
long orbital period results in a maximum orbital smearing of only 1.8\kms.

\begin{figure}
\begin{center}
  \includegraphics[width=5.5cm,height=8.5cm,angle=-90]{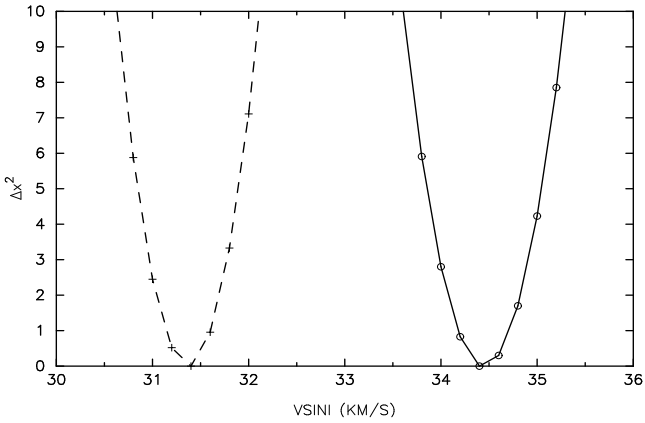}
\end{center}
\caption{The results of the \vsini\ analysis using the green and red spectra. The
variation of \vsini\ with $\chi^2$ is shown for the continuum limb-darkening (solid line) and the zero limb-darkening (dashed line) case.}
\label{chisq}
\end{figure}

The secondary star's rotational broadening combined with the radial velocity
semi-amplitude is \texttt{essential} for determining the binary mass ratio. In principle
one would determine the mass ratio directly by comparing the secondary
star's spectrum with a model spectrum of a Roche-lobe filling secondary star,
\citep{Shahbaz03}.
However, this requires a extremely high quality data (S/N $>$150), 
which our data do not have.
Therefore in order to determine the rotational broadening of the secondary star \vsini\
we follow the standard procedure described by \cite{Marsh94}.  We subtracted a
constant (representing the fraction of light from the template star) multiplied
by a rotationally broadened version of the template star. The optimal
subtraction routine adjusts the constant, minimizing the residual scatter
($\chi^2$) between the spectra. We broadened the template star spectrum from 10
to 90\kms\ in steps of 0.2\kms\ using a spherical rotation profile \citep{Gray92}
with a linear limb-darkening coefficient appropriate for a K0\,III star at the
central wavelengths of the green and red spectra.   The analysis was carried on
the green and red spectra only, because of the poor signal-to-noise in the blue
spectrum. The regions used in the analysis  excluding the HI and HeI  emission
line regions and interstellar features  Na\,I\,5989.95, 5895.92\angs\ and
6284\angs. The additive nature of the $\chi^2$ distribution allows us to add
the  $\chi^2$ distribution for the green and red spectra, which is the same as
performing  the \vsini\ analysis using the combined green and red spectra. We
obtained a  minimum reduced $\chi^2$ of 0.97 corresponding to a \vsini\ of
34.5$\pm$0.3\kms (see Figure\,\ref{chisq}).
Similar results have also been recently obtained by \cite{Steeghs07}.

It should be noted that the analysis above assumes that the limb-darkening
coefficient appropriate for the radiation in the lines is the same as for the
continuum. However, in reality this is not the case, and the absorption lines
in late-type stars will have core limb-darkening coefficients much
smaller than that appropriate for the continuum  \citep{Collins95}. In order to
determine the extreme limits for \vsini\ we also repeated the above analysis
using zero limb-darkening and  obtained a \vsini\ of 31.4$\pm$0.2\kms. As one
can see, the main uncertainty in the  \vsini\ determination comes from the
uncertain limb-darkening coefficient.

What is clearly evident is that the rotational broadening determined
using UVES  spectra is much lower compared to the rotational
broadening of 64$\pm$9\kms\ obtained previously using 
New Technology Telescope (NTT) data
\citep{Shahbaz99}. We discuss this in more detail  in
section~\ref{ntt}.

\section{Using Least-Squares Deconvolution}
\label{LSD}

Ideally what  we would like to do is observe a single high
signal-to-noise  isolated absorption line, from which we can directly
measure the shape of the line profile.   The Least-Squares
Deconvolution (LSD) method \citep{Donati97}  allows us to do exactly
this. LSD is a cross-correlation technique that  effectively stacks
the thousands  of stellar absorption lines in an  echelle spectrum to
produce a single 'average' absorption line  profile with increased
signal-to-noise ratio; theoretically the increase in signal-to-noise
ratio is the square root of the number of lines observed.  It has been
used in the spectropolarimetric observations of active stars
\citep{Donati97} and in Doppler imaging studies  (e.g. see
\citealt{Barnes04}). More recently, LSD has been used in conjunction with
Roche-tomography to map the surface distribution  of the secondary
stars in cataclysmic variables (e.g. \citealt{Watson06}). With LSD we
obtain a single high-quality line profile from which we can readily
determine the  rotational broadening.

LSD assumes that all the absorption lines are rotationally  broadened
by the same amount, and hence just requires the position and strength
of the observed  lines in the echelle spectrum to be known. We
generated a line list  appropriate for a K0\,III star from the Vienna
Atomic Line database (\citealt{Kupka99}; \citealt{Kupka00}). Approximately
3100 lines were used in the deconvolution process across both the red and
green spectra. Our version of LSD propagates the errors through the
deconvolution process.

\begin{figure}
\begin{center}
  \includegraphics[width=5.5cm,height=8.5cm,angle=-90]{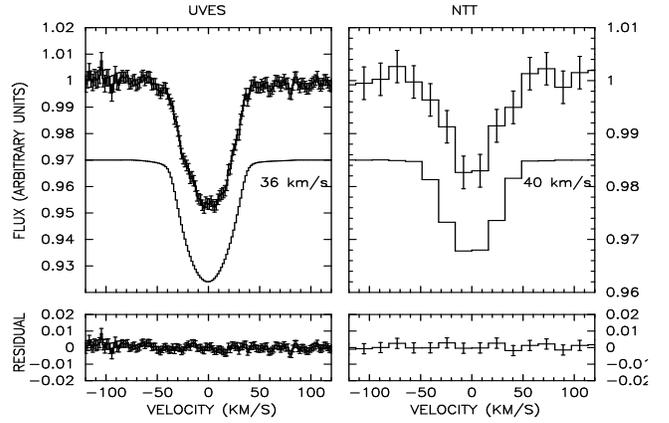}
\end{center}
\caption{The results of the \vsini\ analysis using LSD profiles.  The uppermost plot
shows the LSD profile determined from the green  and red
spectra of \target. The LSD profile below (offset for clarity) shows the LSD
profile of the K0\,III template star rotationally broadened using a continuum
limb-darkened  spherical rotation profile and scaled to match the  top profile.
The bottom panel shows the residual after the optimal subtraction. The left and
right panels show the results for the UVES and NTT data respectively.}
\label{lsd}
\end{figure}

LSD profiles were computed for the individual flux calibrated  \target
and  K0\,III star spectra.  The LSD profiles of \target were then
averaged  and compared to the rotationally broadened version of the
template star LSD profile using the optimal subtraction
procedure. Basically, we repeat the standard method described earlier,
but this time  using the LSD profiles of the green and blue spectra
which only contain a single absorption line (see Figure\,\ref{lsd}).  We
obtain \vsini\ values of 31.3$\pm$0.5\kms\ and  34.7$\pm$0.5\kms\
using zero and continuum limb-darkening respectively for the combined
green and red LSD profiles  (minimum reduced $\chi^2$ of 0.90 and 0.97
and respectively).  Although these values agree well with what we
obtained in section\,\ref{vsini},  there are less systematic
uncertainties in the analysis of the LSD line profiles, primarily
because one can clearly discern the shape of the high signal-to-noise
absorption line profile (see section\,\ref{ntt}).

We can now use the determined \vsini\ (assuming limb-darkening) and optimally
subtract the broadened template spectra from the \target and determine the
fractional contribution of the K0\,III secondary star to the total flux. 
We obtain  0.26$\pm$0.02,
0.30$\pm$0.01 and 0.33$\pm$0.01 for the blue,  green and red spectra
respectively. 

\subsection{Re-analysis of the NTT data using LSD.}
\label{ntt}

The data published in \cite{Shahbaz99} was taken with the NTT+EMMI and 
has a  FWHM resolution of
0.83\angs\ (=38.3\kms). Our analysis of the data revealed a \vsini\ of 
64$\pm$9\kms. [It should be noted that the template star used in the analysis
of the NTT data of \target was actually taken at the WHT in 1992
\citep{Casares94}  and had a resolution of 0.5\angs, better than the NTT data.]
In an attempt to understand why the NTT data gives a higher value for the
secondary star's rotational broadening, we computed the LSD profile of the
average spectrum presented in \cite{Shahbaz99}. We then performed the standard
\vsini\ analysis and optimal subtraction, using the LSD profile of the UVES
K0\,III star degraded  to match the resolution of the NTT data. Using the
appropriate continuum limb-darkening coefficient (as in \citealt{Shahbaz99}),
we obtain a \vsini\ of 40$\pm$10\kms, consistent with the value we obtained
using the UVES data (see Figure\,\ref{lsd}). 
Note that a \vsini\ of 40\kms corresponds to a 
rotational profile with an equivalent FWHM of 60.6\kms, which is only a factor
of 1.6 more than the instrumental resolution.

This difference is most likely due to a combination of the low signal-to-noise
and resolution of the NTT data, which results in the broad absorption blend 
near  6495\angs\  dominating the $\chi^2$ of the comparison between the target
spectrum and the scaled rotationally broadened template star spectrum. We 
tested this hypothesis by performing our original analysis (as was done in
\citealt{Shahbaz99}) but masking the broad 6495\angs\ blend. Indeed we find
that we  obtain a \vsini\ of 41$\pm$15\kms\ compared to 64$\pm$9\kms\ when using
all the features, suggesting that the 6495\angs\ blend adds a systematic
uncertainty in the determination of \vsini\ when using low-resolution low
signal-to-noise data. We thus encourage the use of LSD as a secondary check 
on the \vsini\  measurements in order to avoid such systematics from passing by unnoticed.

\begin{figure}
\begin{center}
  \includegraphics[width=5.5cm,height=8.5cm,angle=-90]{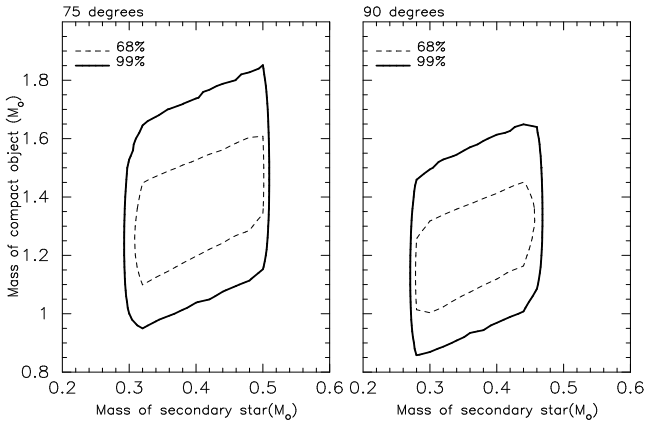}
\end{center}
\caption{The mass of the binary components obtained using a  Monte Carlo
simulation (see section\,\ref{masses}). The solid and dashed lines show the 99\%
and 68\% confidence levels respectively. The left and right panels are for
binary inclination angles of 75$^\circ$ and 90$^\circ$ respectively.}
\label{masses}
\end{figure}

\section{The system parameters}

Since the companion star fills its Roche lobe, \texttt{assuming it is } synchronized with the  
binary motion, the rotational broadening provides a direct measurement of the
binary mass ratio, $q$ (=$M_2$/$M_1$: $M_1$ and $M_2$ are the mass of the compact object 
and secondary star respectively), through the expression 
$v\,\sin i = K_2 (1+q) R_2/a$ 
\citep{Horne86},
where $K_2$, $R_2$ and $a$ are the secondary star's radial velocity semi-amplitude, 
radius and the binary separation, respectively.
Therefore,  substituting the values for $K_2$
\citep{Shahbaz04} and \vsini~we find $q$=0.281$\pm$0.034. Furthermore, using
$K_2$ and $q$, with the orbital period   $P_{\rm orb}$ and the binary
inclination $i$, we can determine the masses of the compact object,  $M_1$, and
the companion star, $M_2$, using the mass function 
$P K_2^3 / 2 \pi G = M_1 \sin^3 i / (1+q)^2$.
In order to determine the uncertainties in $q$, $M_1$ and $M_2$ we used a Monte
Carlo simulation, in which we draw random values for the observed quantities
which follow a given distribution, with mean and variance the same as the
observed values.  For $K_2$ the distribution is taken to be Gaussian as the
uncertainty  is symmetric about the mean value.  However for \vsini~we take a
uniform random distribution,  because of the uncertainties in the 
limb-darkening; 31.3-$2.33\sigma$ \kms\ to 34.7+2.33$\sigma$ \kms\ (99\% confidence).
 
Given the measured masses and orbital period, we use Kepler's Third Law to
determine the semi-major axis $a$. Eggleton's expression for the effective
radius of the  Roche-lobe \citep{Eggleton83} then determines the radius of the
secondary $R_2$, the temperature (4800\,K) is inferred from the spectral type 
\citep{Gray92} and finally  Stefan-Boltzmann's law determines the luminosity \lsol. 
Figure\,\ref{masses} shows the allowed mass range for the binary components  for
the limits on the inclination angles of 75$^\circ$ and 90$^\circ$
(\citealt{Shahbaz04}), and 
Table\,\ref{param} gives the determined system parameters.
Similar results have also been recently obtained by \cite{Steeghs07}.

\begin{table}
\caption{Summary of derived system parameters. The uncertainty in the binary
masses are 1-$\sigma$ and the others are at the 
99\% confidence level.
}
\label{param}
\begin{center}
\begin{tabular}{lcc}\hline \hline
Parameter          & $i$=75$^\circ$    & $i$=90$^\circ$ \\
\hline
$M_1$ ($M{_\odot}$)& 1.37 $\pm$ 0.13   & 1.23 $\pm$ 0.12   \\
$M_2$ ($M{_\odot}$)& 0.30 -- 0.49      & 0.27 -- 0.45	   \\
$R_2$ ($R{_\odot}$)& 5.56 -- 6.61      & 5.37 -- 6.39	   \\
$L_2$ ($L{_\odot}$)& 14.5 -- 23.3      & 13.6 -- 19.2	   \\
$a$   ($R{_\odot}$)& 20.0 -- 20.5      & 19.2 -- 22.5	   \\ \hline	 
\end{tabular}
\end{center}
\end{table}

\section{Interstellar reddening}

\begin{figure}
\begin{center}
  \includegraphics[width=5.5cm,height=6.5cm,angle=-90]{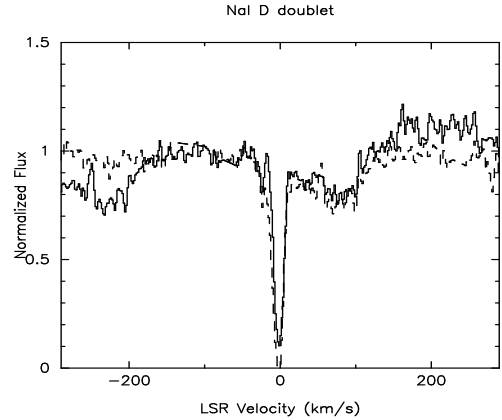}
\end{center}
\caption{The Na\,I D1 (5895.92\angs) and D2 (5889.95\angs) 
doublet converted to the frame of the local standard at rest (LSR). The solid
and dashed lines are the D1 and D2 components respectively.}
\label{NaD}
\end{figure}

We clearly detect the NaI D interstellar lines in absorption which, although 
very deep, are not saturated (see Figure\,\ref{NaD}).
The equivalent
width (EW) of the NaI D1 (5895.92\angs) and D2 
(5889.95\angs) components  were measured to be 0.56$\pm$0.02\angs\ and
0.74$\pm$0.02\angs\ respectively.  
Note that the HeI feature at 5875.62\angs\ is not broad and so does not contaminate the
NaI DI feature (see Figure\,\ref{features}).
The ratio between the two Na lines is much
less than the factor of two expected at the lowest optical depths
\citep{Munari97}. Using the  empirical relationship of \cite{Barbon90} the 
total EW of the doublet corresponds to E(B-V)=0.33$\pm$0.01\,mag. For
comparison, the Hydrogen column as estimated from the Schlegel dust maps
\citep{Schlegel98} gives E(B-V)=0.25 mags, whereas the  hydrogen column
estimated from the HI maps of \cite{Dickey90}  gives E(B-V)=0.31 mags
($N_H=1.69\times10^{21}$\,atoms\,cm$^{-2}$) assuming E(B-V)=$A_V$/3.1;
 and $N_H=1.79\times10^{21}A_V\,$atoms\,cm$^{-2}$ (\citealt{Rieke85}; 
\citealt{Predehl95}).

\section{Discussion}

Comparing \target\,with Cyg\,X--2 one notes some very interesting
similarities.  Both systems are long period binaries in the halo of
the Galaxy, they are at high inclination angles with evolved
secondaries and have similar binary mass ratios, $q$=0.34
\citep{Casares98} and $q$=0.281 (section\,\ref{masses} )for Cyg\,X--2
and \target respectively. Cyg\,X--2 is known to contain a neutron
star because type I X-ray bursts are observed \citep{Kahn84}. Although
no bursts or pulsations have been seen in V395 Car, our mass
determination implies that it contains a neutron star
(section\,\ref{masses}). However, it should be noted that the
secondary star in \target is much cooler (K0\,III) and hence less
luminous than the secondary star in Cyg\,X--2 (A9\,III ;
\citealt{Casares98}), contrary to what we might have expected given
their inferred masses.  This difference is related to the very
different evolutionary histories.  The secondary star in Cyg\,X--2
started to transfer mass near the end of its main-sequence phase
(Case\,AB; \citealt{Pod00}). Most of the mass of the secondary was
ejected from the  system during an earlier rapid mass transfer phase
and currently we observe the hot inner core of what had been initially
a more massive star.

We can compare the mass and radius of a normal K0\,III star ($M_2$=2.3\msol; 
$R_2$=11\rsol; \citealt{Gray92}) to the observed values of \target of $M_2\sim$0.4\msol\
and  $R_2\sim$6\rsol. The observed values suggest that the secondary star in
\target is more evolved and has a lower mass  compared to a normal K0\,III
star. Placing the observed secondary star on the Hertzsprung-Russell  diagram we 
find that its  position corresponds to a normal single star with mass
0.6--1.4\msol\  that has crossed the Hertzsprung gap and now lies on the
Hayashi line.   The evolution of the binary is dominated by the evolution of
the evolved secondary star and the mass transfer is early massive Case\,B,
because such a star no longer burns  hydrogen in its core (Case\,A can be ruled
out because the observed secondary star does not resemble a single  star of
the same mass). For a binary with a secondary star near the onset of such mass
transfer, one requires $q<$1 throughout its history \citep{Kin99}. The position
of a secondary on the  Hertzsprung-Russell diagram undergoing early massive
Case\,B mass transfer is close to that of a single star of the same mass. For a
1.4~\msol\ compact object this implies $M_{\rm 2}<$1.4~\msol, which is consistent
with the single star mass corresponding to the observed properties of the
secondary star.

\begin{figure}
\begin{center}
\includegraphics[width=5.5cm,height=6.5cm,angle=-90]{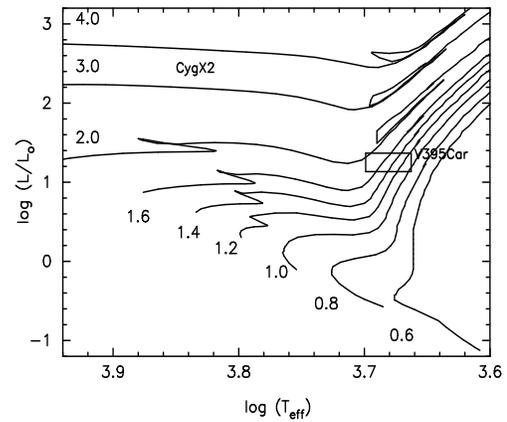}
\end{center}
\caption{Hertzsprung-Russell diagram with evolutionary tracks for solar
metalicity  stars in the range 0.6--3.0\msol \citep{Girardi00}. The box  shows
the position of the  observed effective temperature  (4800$\pm$200\,K) and
luminosity of the secondary star in \target (see Table\,\ref{param}). We also
show the position of Cyg\,X--2 which despite having similar  system parameters,
has a very different evolutionary history compared to V395\,Car.}
\label{HRD}
\end{figure}

\section{Conclusions}

We have accurately measured the rotational broadening (\vsini) in
V395\,Car  using UVES echelle spectroscopy and the method of
Least-Squares Deconvolution (LSD), which provides a single high
signal-to-noise ratio absorption line profile.  We determine \vsini\
to be 31.3$\pm$0.5\kms to 34.7$\pm$0.5\kms (assuming zero and
continuum limb darkening, respectively). The disagreement with
previous results is most likely due to a combination of low
signal-to-noise ratio  and the relatively low spectral resolution of
the data.  We find that using relatively low-resolution spectroscopy
can result in systemic uncertainties in the  measured \vsini\ values
obtained using standard methods. We suggest the use of LSD as a
secondary,  reliable check of the results because LSD allows one to
directly discern the shape of the absorption line profile.  Using our
revised \vsini\ value in combination with the secondary star's radial
velocity gives a binary mass ratio of 0.281$\pm$0.034. Furthermore,
assuming a binary inclination angle of 75$^\circ$ gives a compact
object mass of 1.37$\pm$0.13\msol, consistent with a canonical
neutron star mass.

\begin{acknowledgements}
TS acknowledges support from the Spanish Ministry of Science  and Technology 
under the programme Ram\'{o}n y Cajal. 
CAW is supported by a PPARC postdoctoral fellowship.
\end{acknowledgements}

\bibliography{aa}

\end{document}